\documentclass[twocolumn,multicolumn,aps,prb,showpacs]{revtex4}
\usepackage{graphicx}
\usepackage{dcolumn}
\usepackage{bm}
\usepackage{color}
\usepackage{latexsym}
\usepackage[T1]{fontenc}
\usepackage{amsmath}

\input{epsf}

\begin{document}

\preprint{APS/123-QED}

\date{\today}

\title{Strong pinning and vortex energy distributions in single crystalline Ba(Fe$_{1-x}$Co$_{x}$)$_{2}$As$_{2}$ }
\author{S. Demirdi\c{s} and C.J. van der Beek}
\affiliation{Laboratoire des Solides Irradi\'{e}s, CNRS UMR 7642 \& CEA-DSM-IRAMIS, Ecole Polytechnique, F91128 Palaiseau cedex, France }
\author{Y. Fasano, N.R. Cejas Bolecek, and H. Pastoriza}
\affiliation{Laboratorio de Bajas Temperaturas, Centro At\'{o}mico
Bariloche \& Instituto Balseiro, Avenida Bustillo 9500, 8400
Bariloche, Argentina}
\author{D. Colson and F. Rullier-Albenque}
\affiliation{Service de Physique de l'Etat Condens\'{e},  L'Orme des M\'{e}risiers, CEA-DSM-IRAMIS, F91198 Gif-sur-Yvette, France }

\begin{abstract}
The interrelation between heterogeneity and flux pinning is studied in Ba(Fe$_{1-x}$Co$_{x}$)$_{2}$As$_{2}$ single crystals with widely varying Co-content $x$.  Magnetic Bitter decoration of the superconducting vortex ensemble in crystals with $x=0.075$ and $x=0.1$ reveal highly disordered vortex structures. The width of the Meissner belt observed at the edges of  the crystals, and above the surface steps formed by cleaving, as well as the width of the intervortex distance distribution, indicate that the observed vortex ensemble is established at  a temperature just below the critical temperature $T_{c}$. The vortex interaction energy and pinning force distributions extracted from the images strongly suggest that the vortex lattice disorder is attributable to strong pinning due to spatial fluctuations of  $T_{c}$ and of the superfluid density. Correlating the results with the critical current density yields a typical length scale of the relevant disorder of 40 - 60 nm.
\end{abstract}

\maketitle

\section{Introduction}

Recent vortex imaging studies of iron-based superconductors have
unveiled highly disordered vortex structures that challenge the
use of traditional analysis  procedures based on the
characterization of positional and orientational lattice
 correlations.\cite{Eskildsen,Eskildsen2009b,Inosov,Vinnikov,Kalisky,Luan2010,Yi Yin}
For example, the combination of small-angle neutron scattering
experiments with Bitter decoration \cite{Eskildsen,Eskildsen2009b}
and Magnetic Force Microscopy \cite{Inosov} revealed a
``vitreous'' phase in Ba(Fe$_{1-x}$Co$_{x}$)$_{2}$As$_{2}$
single-crystals. The latter work shows that the vortex structure
of the overdoped material ($x=0.19$) presents, at best,
short-range hexagonal order in the field range of $10^{-3}$ to
9\,T. Disordered vortex structures were also observed by means of
Bitter decoration in single-crystals of other iron-based pnictide
superconductors including Ba$_{1-x}$K$_{x}$Fe$_{2}$As$_{2}$,
Sr$_{1-x}$K$_{x}$Fe$_{2}$As$_{2}$, and SmAsO$_{1-x}$F$_{x}$.
\cite{Vinnikov} Regardless of material, doping, and synthesis
method, the disordered vortex structures are attributed to a
strong pinning the nature of which was not discussed.

The only reported ordering effect on the orientation of the vortex ensemble is that
induced by twin-boundaries in Ba(Fe$_{0.949}$Co$_{0.051}$)$_{2}$As$_{2}$. \cite{Kalisky}
This scanning Superconducting Quantum Interference Device (scanning SQUID) microscopy study shows that vortices  avoid twin boundaries acting as a barrier for vortex motion. \cite{Kalisky} These results echo
earlier work on the doping-dependence of the critical current density $j_{c}$,  that
suggests that structural domain walls may act as effective pinning centers in the
underdoped material.\cite{Prozorov2}

Concerning the nature of the strong pinning ubiquitous to
iron-based superconductors, inhomogeneities in the dopant ions
distribution was suggested to be at the origin of a dense vortex
pinning nanostructure in the case of
Ba(Fe$_{1-x}$Co$_{x}$)$_{2}$As$_{2}$ ($x=0.1$). \cite{Yamamoto}
The same study shows that since thermal fluctuations are weak, the
finite width of the superconducting transition can only be
ascribed to an inhomogeneous $T_{c}$ distribution due to local
compositional variations. \cite{Yamamoto} Furthermore, scanning
tunnelling spectroscopy studies in different iron-based pnictides
reveal nanoscale variations of the local superconducting
gap.\cite{Yi Yin 2009b, Massee2009, Fasano2010} In
 Ba(Fe$_{1-x}$Co$_{x}$)$_{2}$As$_{2}$,  the length scale on which
the deviations from the average gap value occur is comparable to the average distance between
dopant ions.\cite{Massee2009} Nevertheless, no correlation between the vortex positions and the
superconducting-gap inhomogeneities or other defects has, as yet, been found. \cite{Yi Yin}

Hence, all techniques agree on the absence of an ordered vortex
structure in iron-based superconductors. However,  there is no
clear consensus on the origin of the disorder in the vortex
ensemble and the pinning causing it.  The aim of this paper is the
characterization of this strong pinning by means of a quantitative
analysis of the spatial distributions of pinning energy and
pinning force. We found that the key to understand the disordered
vortex configurations is that these are frozen at $T  \sim T_{c}$,
in crystals with important spatial variations of the
superconducting parameters. The correlation of the extracted
pinning forces and energies with measurements of $j_{c}$ indicates
that spatial variations of the superfluid density and of $T_{c}$ ,
on
 the scale of several to several dozen nm, are the most relevant for pinning.

\begin{figure}[t]
\includegraphics[width=0.37\textwidth]{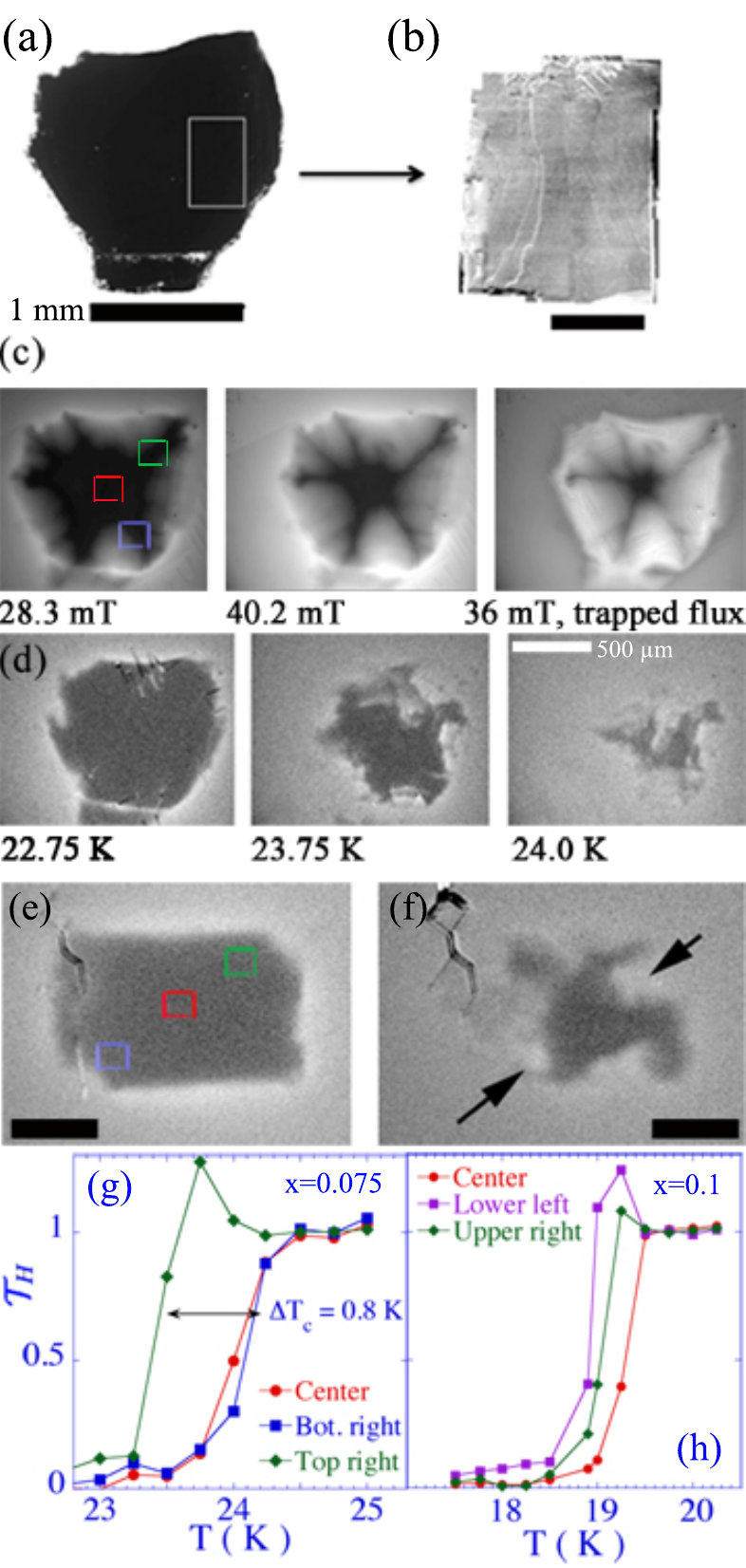}%

\caption{(Color online) (a) Photograph of
Ba(Fe$_{0.925}$Co$_{0.075}$)$_{2}$As$_{2}$ crystal \# 2. (b)
Scanning electron micrograph of the decorated sample \#2.1 cut
from the larger crystal   \#2. (c) Magneto-optical images of Ba(Fe$_{0.925}$Co$_{0.075}$)$_{2}$As$_{2}$ crystal \#2 at $T=15$\,K and indicated values of the applied magnetic field. (d) Differential Magneto-Optical (DMO) images in the vicinity of $T_{c}$ for $\mu_{0}\Delta H_{a}=0.1$\,mT. (e) DMO image of Ba(Fe$_{0.9}$Co$_{0.1}$)$_{2}$As$_{2}$ crystal \#1 at  full screening ($T = 17.5$\,K) and (f) at mid-transition ($T = 19.25$\,K). The arrows indicate regions of paramagnetic transmittivity at the superconducting transition.  (g) Local transmittivity $\mathcal{T}_{H}$ measured on the three regions of 
 Ba(Fe$_{0.925}$Co$_{0.075}$)$_{2}$As$_{2}$  crystal \# 2. (h) $\mathcal{T}_{H}$ measured on
 the three regions of Ba(Fe$_{0.9}$Co$_{0.1}$)$_{2}$As$_{2}$ crystal \#1 indicated
 in (e). Scale bars correspond to a length of 100 $\mu$m unless indicated otherwise.}

\label{fig:mo}

\end{figure}

\section{Experimental details}

Single-crystals of Ba(Fe$_{1-x}$Co$_{x}$)$_{2}$As$_{2}$  were
grown using the self-flux method.\cite{Florence} Starting reagents
of high-purity Ba, FeAs and CoAs were mixed in the molar ratio
1:(4-x):x, loaded in alumina crucibles and then sealed in
evacuated quartz tubes. For each doping level, chemical analysis
by an electron probe was performed on several crystals yielding
the Co content within 0.5\% absolute accuracy. For this work we
studied six doping levels. 

The penetration of  magnetic flux into selected crystals of
thickness 30\,$\mu$m was visualized by the magneto-optical imaging
(MOI) method.\cite{Dorosinskii92} A ferrimagnetic garnet indicator
film with in-plane anisotropy is placed on top of the sample and a
polarized light microscope is used to observe it. The Faraday
rotation of the indicator allows the detection of regions with
non-zero perpendicular component of the magnetic flux density
$B_{\perp}$, revealed as bright when observed through an analyzing
polarizer. Dark regions correspond to  $B_{\perp} \approx 0$. In
order to characterize the inhomogeneity of the crystals in the
vicinity of the critical temperature we use the differential
magneto-optical method (DMO).\cite{Kees1} Images acquired at
applied fields $H_{a}+ \Delta H_{a}$ and $H_{a}$ are subtracted,
and the differential images averaged by repeating the procedure 50
times. In the present experiments $\mu_{0}\Delta H_{a}=0.1$\,mT (with $\mu_{0} \equiv 4\pi \times 10^{-7}$ Hm$^{-1}$).

The field dependence of the global critical current density of selected crystals
was obtained from magnetization-loop measurements conducted
 using a Quantum Design SQUID magnetometer. The critical current densities were extracted using the
Bean-critical state model. As discussed below, the assumption of
this model is justified by the way flux penetrates into the
crystals.  Within the Bean model, $j_{c} = 3 \mathcal M / V a$, where
$\mathcal M$ is the magnetic moment, $V$ is the sample volume, and
$2a$ the sample width.\cite{Brandt98}

\begin{figure}[t]
\includegraphics[width=0.4\textwidth]{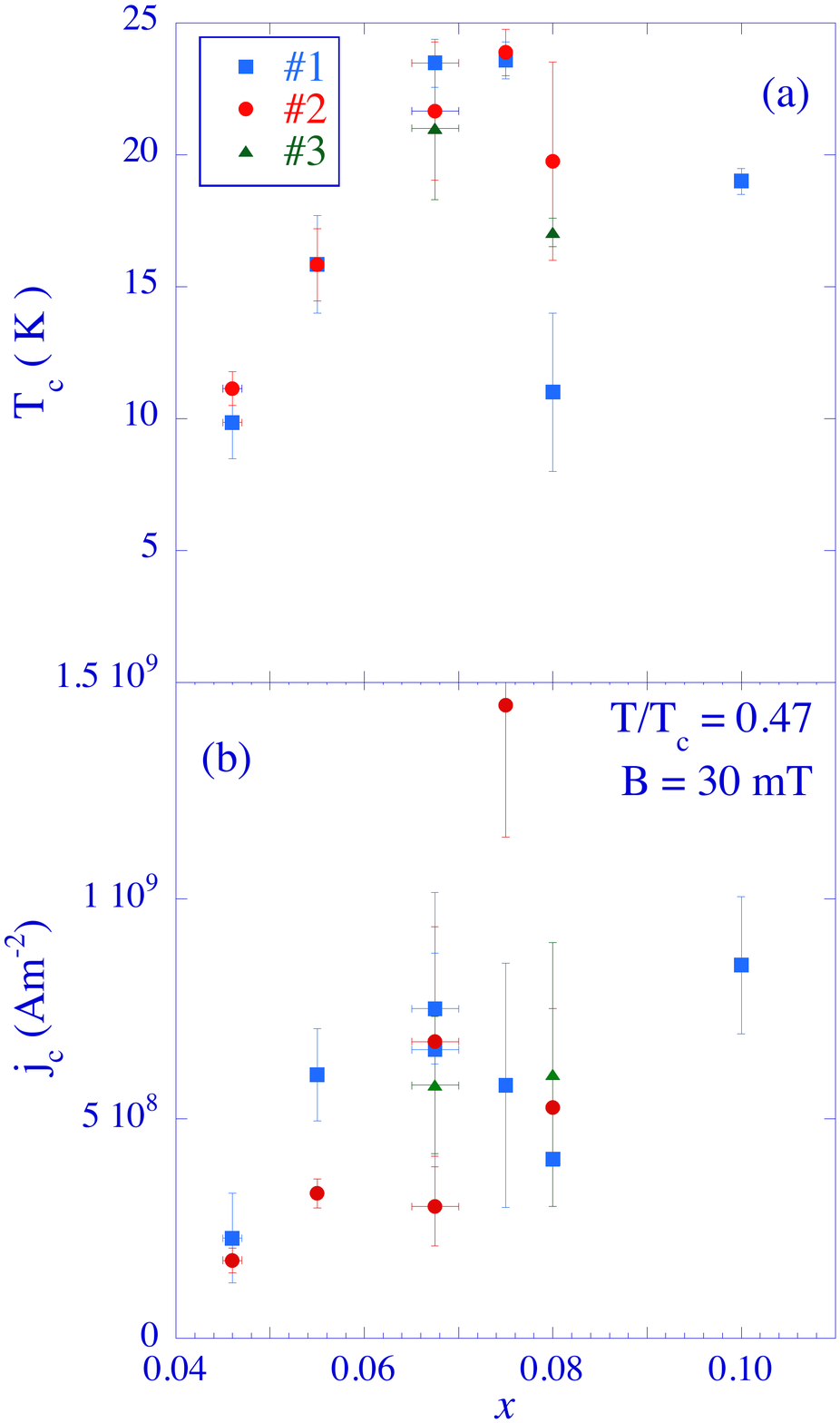}
\caption{ (Color online) (a) Transition temperature,
 $T_{c}$, versus Co doping-level. The error bars denote the local spread
 of $T_{c}$ values within a given crystal. For each doping level, \#1, \#2  and \#3
 denote different crystals. For $x = 0.075$ and $0.1$, the numbering denotes that of
 the decorated crystals. (b) Co doping-level dependence of the critical current density
 $j_{c}$  measured by MOI at  $B=30$\,mT  and a reduced
 temperature of $T/T_{c}=0.47$.}
\label{fig:Tc}
\end{figure}


For the Bitter decoration experiments,\cite{Fasano2008} rectangles of dimension  200\,$\mu$m
$\times$ 300$\,\mu$m were cut from larger crystals using a 20\,$\mu$m wire saw and
1\,$\mu$m SiC grit. Bitter decorations were only performed on
crystals with $x= 0.055$, $x=0.075$, and $x=0.1$. The sample surfaces were freshly cleaved
before the experiments (Fig.\,~\ref{fig:mo}a, b). The 
experiments  were carried out at liquid Helium temperature
($4.2$\,K) and He-exchange gas at pressures of the order of
$200$\,mTorr. The images shown here are the result of
field-cooling experiments at a field $\mu_{0}H_{a} =1$\,mT applied
parallel to the $c$-axis of the crystals. The decorated vortex
arrangement was observed by scanning electron microscopy at
room-temperature.

\section{Results}

\subsection{Magneto-Optical Imaging and $j_{c}$ measurements}

Figure\,~\ref {fig:mo}\,(c) shows  examples of magneto-optical images, here obtained
at $T=15$ K on single-crystal \#2 of the composition with $x=0.075$. The images reveal a
globally  homogeneous penetration of the magnetic flux into the
sample obeying the Bean critical state.~\cite{Bean,Zeldov} We
obtain the local value of the critical current density  from
$j_{c} \sim 6~\partial B_{\perp} /\partial x$ (the factor 6 is
estimated from Ref.~\onlinecite{Brandt98} for a crystal
aspect-ratio of 0.1). The DMO images in Fig.\,~\ref
{fig:mo}\,(d) reveal the same Bean-like flux penetration with an
inhomogeneous $j_{c}$ arising from the spatial variation of
$T_{c}$.

This inhomogeneity can be quantified using a plot of the local transmittivity,
 defined as the ratio 
$\mathcal{T}_{H}=[I(\mathbf r,T) - I(\mathbf r,T \ll
T_{c})]/[I(\mathbf r,T \gg T_{c})-I(\mathbf r,T \ll T_{c})]$ of
the relative local luminous intensities $I(\mathbf r,T)$ in the
DMO images. The temperature-dependence of $\mathcal{T}_{H}$
measured on different regions of crystals \#2 and \#1 is depicted
in Figs\,.~\ref {fig:mo}\,(e),(f). The local variation of
$T_{c}$-values within a given crystal is of the order of 0.5 -- 1
K. In addition, regions of lower $T_{c}$ give rise to a
paramagnetic signal at the transition due to flux concentration by
the surrounding superconducting parts of the crystal.

Figure\,~\ref{fig:Tc}\,(a) summarizes the width of the 
$T_{c}$ distribution for a large number of
Ba(Fe$_{1-x}$Co$_{x}$)$_{2}$As$_{2}$ single-crystals  of different
doping levels. Figure\,~\ref{fig:Tc}\,(b) shows the Co
doping-level dependence of $j_{c}$  for the same series of
single-crystals at a reduced temperature of $T/T_{c}=0.47$. A
rather large sample-to-sample variation of the low--field
($B_{\perp} = 30$\,mT) $j_{c}$ is observed. Certainly, no clear
doping-dependent trend appears, as proposed in
Ref.\,~[\onlinecite{Prozorov2}]. The obtained critical-current
values are comparable to those reported in the literature for the
same material.\cite{Prozorov2008}

\subsection{Vortex imaging}

\begin{figure}[t]
\includegraphics[width=0.47\textwidth]{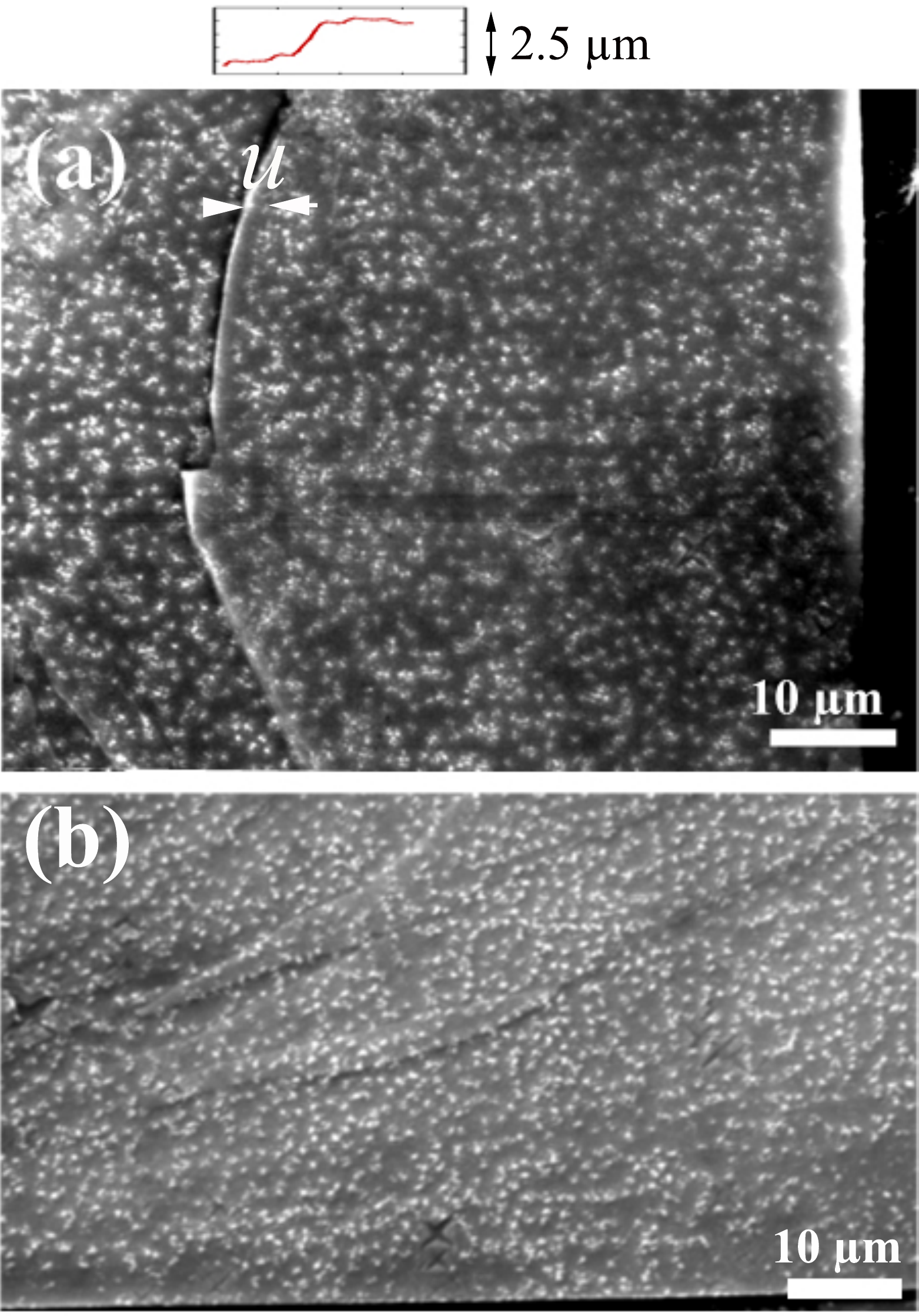}

\caption{Bitter decoration images of
Ba(Fe$_{1-x}$Co$_{x}$)$_{2}$As$_{2}$ single-crystals (a) crystal
\#2.1 with $x=0.075$, and  (b) crystal \#1 with $x=0.1$. The graph above panel (a) shows a profilometer measurement upon crossing the step that traverses the image from top to bottom; full vertical scale is 2.5\,$\mu$m. The width $u$ of the Meissner belt behind the step is also indicated in (a).}
\label{fig:deco}
\end{figure}

The Bitter decoration technique\cite{Fasano2008} was used to observe vortex structures on three of the crystals used to compile Fig.~\ref{fig:Tc}, more precisely, on crystal  \#1 of the composition with $x=0.1$,  crystal \#2 with $x=0.075$, and on crystal \#2 with $x= 0.055$. The decoration of crystal \#2 with $x = 0.055$ was unsuccessful, presumably due to the large value of the penetration depth at low doping. The decorated patterns reveal highly-disordered vortex structures
as in Refs.\,~[\onlinecite{Vinnikov,Eskildsen,Inosov,Kalisky,Luan2010}].
Figures\,~\ref{fig:deco}\,(a) and (b)  reveal regions of high and
low vortex density, as well as the formation of vortex-free zones
near the crystals edges and surface steps, due to the circulating
Meissner current. These images are representative of those
obtained on other regions of the crystal surfaces after different
cleavage runs, and on other crystals. From the images, we extract
the average value of the magnetic induction  as  $B_{int} =
n_{v}\Phi_{0}$, where $n_{v}$ is the vortex density  and
$\Phi_{0}=h/2e$ is the flux quantum. For all images we obtain an
average induction $B_{int}\approx 0.8$\,mT, 20\% smaller than the
applied field $H_{a}=1$\,mT.

\begin{figure}[t]
\vspace{9mm}
\includegraphics[width=0.42\textwidth]{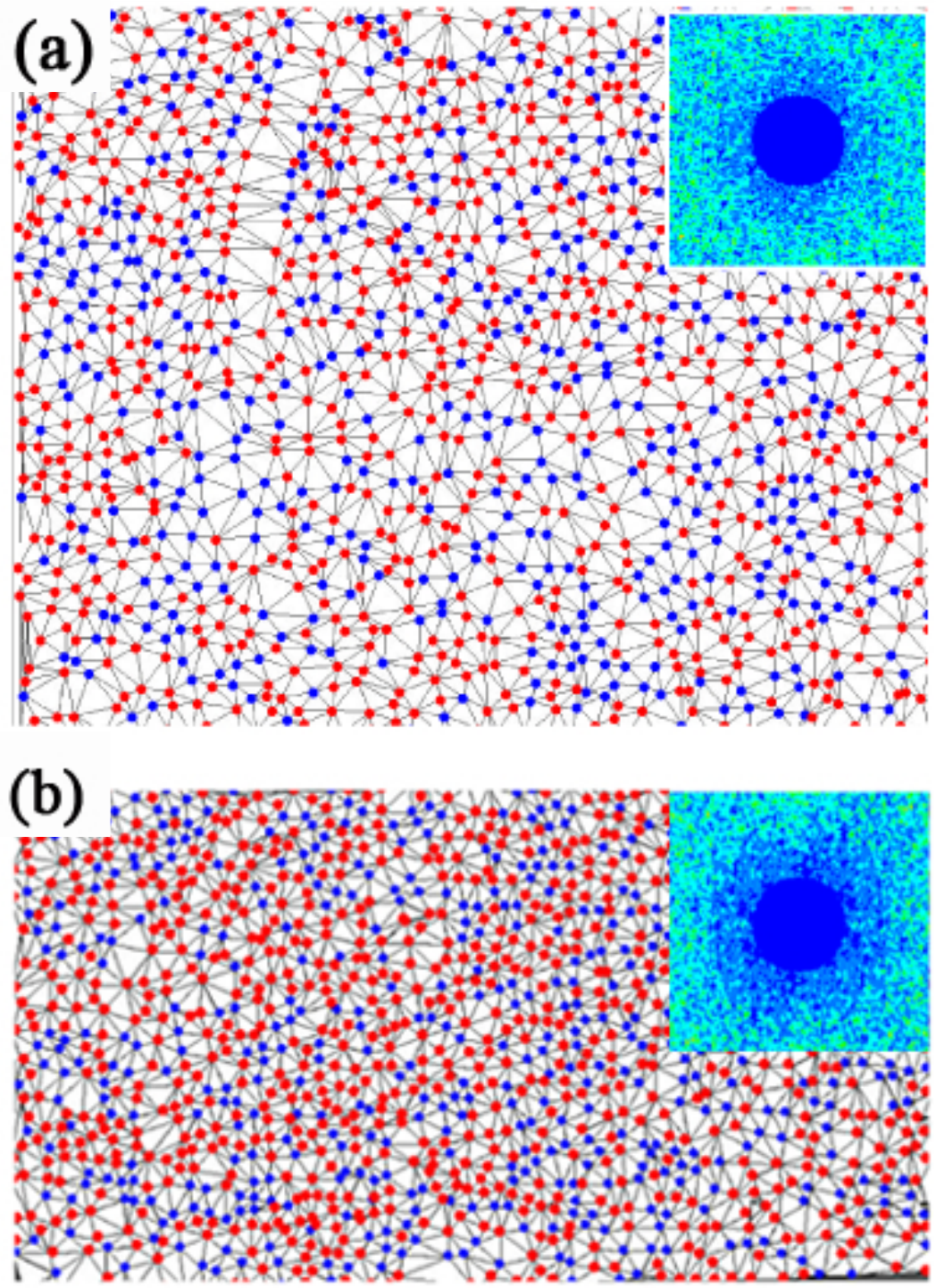}
\caption{(Color online) Delaunay triangulation of the vortex
structure of Ba(Fe$_{1-x}$Co$_{x}$)$_{2}$As$_{2}$ single-crystals
 (a)\# 2.1 with $x=0.075$ and (b) \#2 with $x=0.1$.
The blue dots represent vortices with 6 nearest neighbors while
red ones represent vortices with a different coordination number.
The insets show the respective Fourier transforms of the vortex
positions.  } \label{fig:tri}
\end{figure}

Figures\,~\ref{fig:tri}\,(a) and (b) present the Delaunay
triangulations of the images in Fig.\,~\ref{fig:deco} for
$x=0.075$ and $x= 0.1$, respectively.  Here,  the blue dots represent
vortices with sixfold coordination while the red dots represent
vortices which have a different coordination number. The insets to
Figs.\,~\ref{fig:tri}\,(a) and (b) show the Fourier transforms of
the vortex positions which once again demonstrate the absence of
any order in the vortex structure.

\subsection{Vortex configurations near surface steps}
\label{sec:3}

The correct determination of the distribution of vortex pinning energies in the crystal and its
interpretation requires knowledge of  the temperature at which the
vortex ensemble was frozen in the observed configuration. To
determine this, we analyze the vortex distribution near the
ubiquitous steps seen on the surfaces of the crystals. Such steps
result from the repeated crystal cleavage performed during the
Bitter decoration experiments. In zero-field cooled experiments,
steps act as obstacles for vortex entry into the sample; they were
described in Ref.~\onlinecite{Pardo} as ``vortex diodes''.

However, the present decoration experiments are  in field-cooled
conditions and hence vortices nucleate in the sample at the same
temperature that the mixed state is stable. As one cools down, the
Meissner screening current running along the crystal edges, but
also along the surface steps, increases as the penetration depth
$\lambda_{ab}$ for currents running in the $ab$--plane decreases.
Thus, while cooling, vortices on the high-side of the step are
progressively repelled by the increasing Meissner current density
$j_{M} \sim H_{a}/\lambda_{ab}$.  At the same time, the proximity
of the step surface results in an attractive force that can be
described by an image vortex segment. Finally, the vortex lattice
elasticity tends to restore a homogeneous flux distribution near
the step. The situation is therefore similar to vortex entry or
exit over a surface barrier.

\begin{figure}[b]
\includegraphics[width=0.30\textwidth]{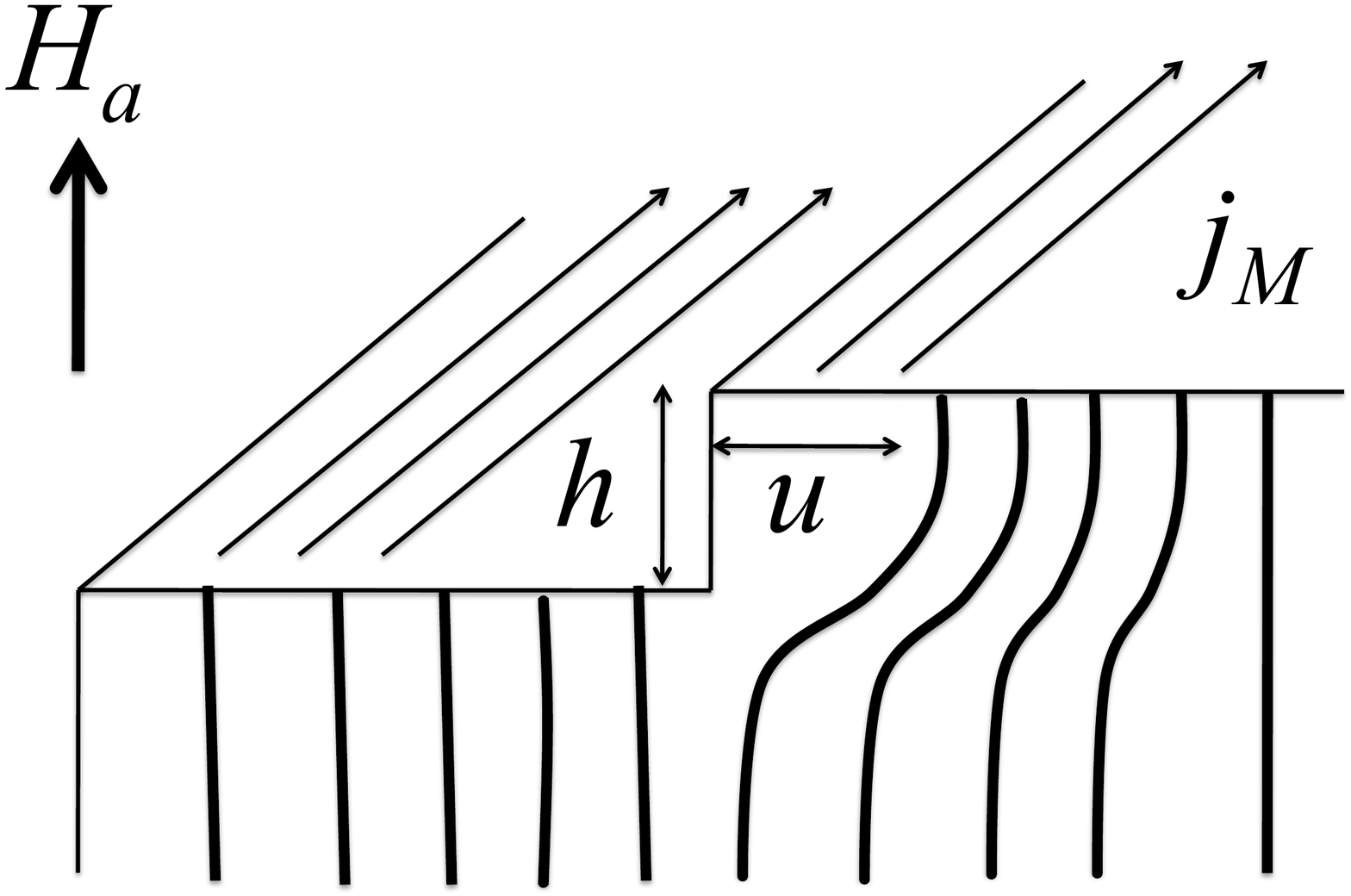}
\caption{Representation of vortex lines near a surface step under 
field-cooled conditions} \label{fig:step}
\end{figure}

At the low fields of interest, the single vortex part of the tilt modulus dominates  vortex elasticity,\cite{Brandt1990} so that  the force balance is
\begin{eqnarray}
\frac{B_{int}}{\lambda_{ab}} \left[ H_{a} {\mathrm e}^{-\upsilon } - \frac{B_{int}}{\mu_{0}}  {\mathrm e}^{-2\upsilon } - \frac{\varepsilon_{\lambda}^{2}\varepsilon_{0}}{\Phi_{0}} \ln \left( \frac{B_{c2}}{2B_{int}} \right) \frac{u\lambda_{ab}}{h^{2}} \right]  & = & \nonumber \\
\frac{B_{int}}{\lambda_{ab}} \left[ H_{a}  {\mathrm e}^{-\upsilon } - \frac{B_{int}}{\mu_{0}}  {\mathrm e}^{-2\upsilon } - \frac{\varepsilon_{\lambda}^{2}\Phi_{0}\upsilon}{4\pi\mu_{0}h^{2}} \ln\left(\frac{B_{c2}}{2B_{int}}\right)\right]&=& 0 .
\nonumber \\ & &
\label{eq:free-energy}
\end{eqnarray}

Here, $\varepsilon_{1} = \varepsilon_{\lambda}^{2}\varepsilon_{0}$
is the vortex line tension, $\varepsilon_{0} =
\Phi_{0}^{2}/4\pi\mu_{0}\lambda_{ab}^{2}$ is the vortex line energy, $ \varepsilon_{\lambda}
= \lambda_{ab}/\lambda_{c}$ is the penetration depth anisotropy,
$\upsilon \equiv u/\lambda_{ab}$, represents the width of the
vortex-free zone behind the step, $u$, normalized to
$\lambda_{ab}$. The step running through Fig.\,\ref{fig:deco}\,(a)
has a height $h = 1.5\,\mu$m while the vortex-free region behind
it has a width $u = 1.8\,\mu$m. Estimating the penetration depth
anisotropy $\varepsilon_{\lambda} \approx 0.16$ from
Refs.\,~\onlinecite{Prozorov2009} and \onlinecite{Hanisch2010},
and with all other parameters known, Eq.\,~\ref{eq:free-energy}
can be solved graphically to yield $\upsilon \sim 1.5$, that is,
$\lambda_{ab} \sim 0.6 u \sim 1.2\,\mu$m. Combining
$\lambda_{ab}$-data from Refs.\,~\onlinecite{Luan2010} and
\onlinecite{Prozorov2009}, we conclude that the observed vortex
pattern is frozen at $T_{f}\approx 0.9 T_{c}$.

\subsection{Pinning energies}

\begin{figure}[tb]
\includegraphics[width=0.50\textwidth]{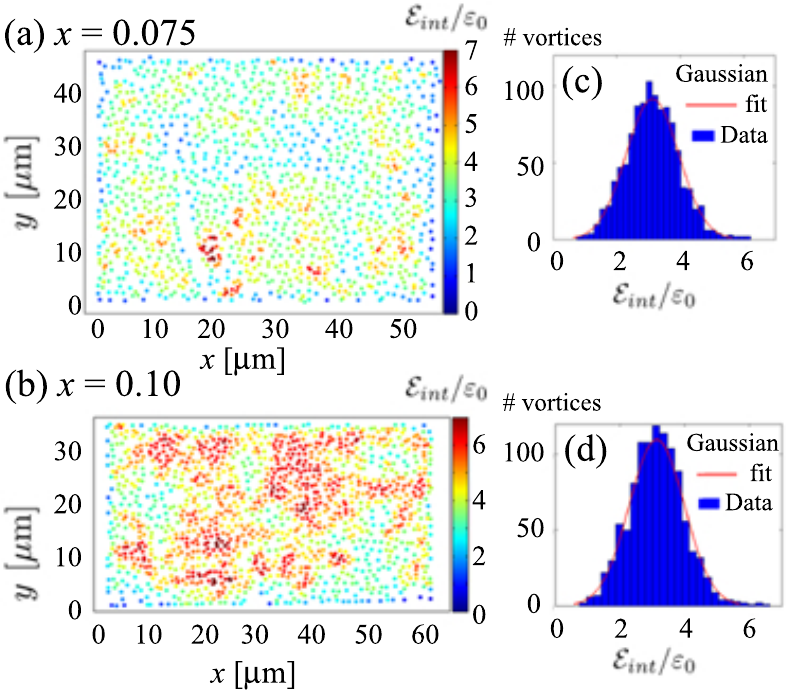}
\caption{(Color online) Left-hand panels: Normalized color-coded
maps of the vortex interaction energy calculated from the images
of Fig.\,~\protect\ref{fig:deco} in
Ba(Fe$_{1-x}$Co$_{x}$)$_{2}$As$_{2}$  single-crystals with (a)
$x=0.075$, and (b) $x=0.1$. Right-hand panels:  Histograms of the
normalized interaction-energy histograms  for (c) $x=0.075$,
and (d) $x=0.1$.} \label{fig:Fig5-energydistribution}
\end{figure}
\begin{figure}[t]
\includegraphics[width=0.5\textwidth]{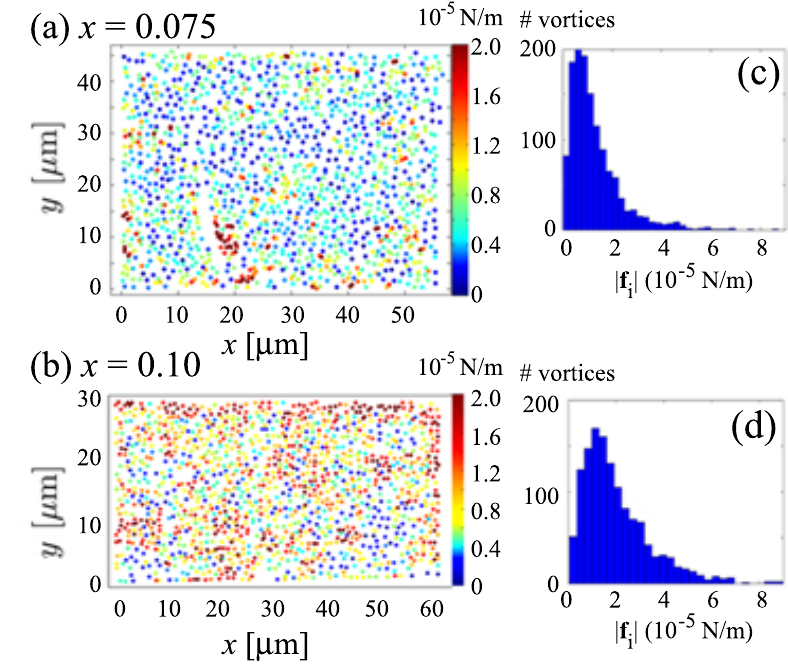}
\caption{(Color online) Color-coded maps of the modulus of the
individual vortex pinning force per unit length, calculated from
the images in Fig.\,~\protect\ref{fig:deco} for
Ba(Fe$_{1-x}$Co$_{x}$)$_{2}$As$_{2}$  single-crystals with  (a)
$x=0.075$ and (b) $x=0.1$. (c) and (d) represent the pinning force distribution for $x=0.075$ and $x=0.1$ respectively.} \label{fig:Fig6-forcedistribution}
\end{figure}

 The inter-vortex interaction energy is  calculated from the vortex positions
 obtained from the decoration images. We calculate the  interaction
 energy
 \begin{equation}
\mathcal E_{int}^{i} = \sum_{j} 2\varepsilon_{0} K_{0}\left(\frac{|r_{ij}|}{\lambda_{ab}}\right)
\label{Eint}
\end{equation}
per unit length along the vortices' direction. $K_{0}(x)$ is the
lowest-order modified Bessel function, and the vortex line energy $\varepsilon_{0} \propto \lambda_{ab}^{-2}$ 
is proportional to the superfluid density. We take into account all vortices $j$
situated at a distance smaller than $10\lambda_{ab}$ from vortex
$i$. This cutoff radius was chosen after verifying that the
interaction energy does not change significantly if greater values
of $j$ are considered. For the determination of the energy
distribution histograms, we only take into account vortices
situated away from the edges of images, at a distance larger than
$4\lambda_{ab}$.\cite{check} We used in this calculation the
penetration-depth value at the temperature at which the vortex
structure was frozen, $\lambda_{ab}(T/T_{c} = 0.9)$ (see
Section\,~\ref{sec:3}).

A similar procedure yields maps of the pinning force acting on an
individual $i$ vortex per unit length
\begin{equation}
\mathbf{f}_{i} = \sum_{j} \frac{2\varepsilon_0}{\lambda_{ab}} \frac{\mathbf{r}_{ij}}{|\mathbf{r}_{ij}|} K_{1}\left(\frac{|\mathbf{r}_{ij}|}{\lambda_{ab}}\right),
\label{fp}
\end{equation}
\noindent  with $K_{1}(x)$ the first-order modified Bessel
function. Since the
 system is stationary, Newton's third law requires the repulsive force
 exerted by neighbor vortices be balanced by the pinning force. A map of the modulus
 $|\mathbf{f}_{i}|$ thus represents a map of the minimum pinning force acting
 on each vortex. In the case of a perfect lattice resulting from negligible effect of pinning,
 the sum (\ref{fp}) vanishes.

We present our results by color-coded maps spanning the whole
decoration images of Figs.\,\ref{fig:deco}\,(a) and (b), and by
histograms of the interaction energy distribution. The interaction
energy maps with the energy-scale normalized by $\varepsilon_{0}$
are shown in Figs.\,~\ref{fig:Fig5-energydistribution}\,(a) and
(b).
A granular structure of denser regions with larger interaction
energy, and dilute regions with smaller $\mathcal E_{int}$ is
clearly visible. This granularity is translated into broad vortex
interaction-energy histograms as shown in
Figs.\,~\ref{fig:Fig5-energydistribution}\,(c) and (d). The
histograms are reasonably well fitted by a Gaussian distribution.
The standard deviations of these histograms are of the order of
23 \%, in contrast with 50 \% for the rather regular vortex structures\cite{BSCCO}  of the same density 
imaged in the high-$T_{c}$ material Bi$_{2}$Sr$_{2}$CaCu$_{2}$O$_{8 + \delta}$.
However, as a result of the high reduced temperature $T_{f}/T_{c}$ at which the vortex ensemble is frozen, the mean interaction energy (normalized by $\varepsilon_{0}$) is ten times larger in Ba(Fe$_{1-x}$Co$_{x}$)$_{2}$As$_{2}$ than in Bi$_{2}$Sr$_{2}$CaCu$_{2}$O$_{8 + \delta}$.\cite{distributions}

The reduced temperature $T_{f}/T_{c}$ at which the vortex ensemble is frozen 
not only affects the deduced interaction energies, but also has a profound effect on the (orientational) order observed in the decorated vortex ensemble.\cite{Pardo97} Pardo {\em et al.}  reported\cite{Pardo97} that in optimally doped Tl$_{2}$Ba$_{2}$CuO$_{6-\delta}$ superconductors with a broad magnetically reversible regime in the temperature-field phase diagram, and concomitantly low $T_{f}/T_{c}$, Bitter decoration yields a regular triangular lattice,  while decorated vortex ensembles in the overdoped material with a narrow reversible temperature range (and high $T_{f}/T_{c}$) are amorphous. 
At the origin of this effect is the high mobility of vortices just above $T_{f}$ in materials with a wide reversible regime, such as 
Bi$_{2}$Sr$_{2}$CaCu$_{2}$O$_{8 + \delta}$ or optimally doped Tl$_{2}$Ba$_{2}$CuO$_{6-\delta}$. On the other hand, the low 
mobility of the vortices just above $T_{f}$ due to strong pinning in the vortex liquid phase in materials [such as, apparently, Ba(Fe$_{1-x}$Co$_{x}$)$_{2}$As$_{2}$] that have a narrow reversible regime yields an amorphous vortex ensemble.

Fig.\,~\ref{fig:Fig6-forcedistribution} shows maps of the modulus
of the pinning force of individual vortices per unit length of the
vortex lines. The pinning energy shows some correlation with the
interaction energy at the local scale: regions of large (small)
$\mathcal E_{int}$ generally correspond to regions of large (small) $|\mathbf
f_{i}|$. There is noticeable inhomogeneity on scales smaller than the apparent grain size.
The juxtaposition of a region with homogeneous large superfluid
density (\em i.e. \rm $\varepsilon_{0}$) with a region of
homogeneous small $\varepsilon_{0}$ would give rise to a larger
pinning force at the interface only. In the images,
 fluctuations of the pinning force within grains of similar  $\mathcal E_{int}$ are observable. Therefore,
inhomogeneity of the superconducting parameters exists not only on
the $\mu$m scale of the images, but also on smaller length scales.

It is interesting to note that the rendered pinning forces are simply related
 to a metastable current density $\mathbf j_{i}$, {\em running through each
 vortex}, as $\mathbf{f}_{i} = (\Phi_{0}/|\mathbf B|) \mathbf B \times \mathbf j_{i}$.
 The average pinning force per unit length of $5 \times 10^{-6}$\,N/m, with local
 maxima of up to $6\times 10^{-5}$\,N/m, imply local currents of the order of
 $2.5 \times 10^{9}$\,A\,m$^{-2}$.  Maximum currents are of the order $3\times10^{10}$\, A\,m$^{-2}$,
  comparable to the low-temperature value of the critical
 current density.

\section{Discussion}
\label{section:discussion}

Since the vortex locations result  from the balance between
inter-vortex repulsion and the interaction $\mathcal E_{p}$ of individual vortices with the
pinning impurities, one has, at  $T_{f}$,  $\mathcal E_{int}= \mathcal E_{p}$. The position of the maximum and the width of the interaction-energy distributions
[see Figs.\,~\ref{fig:Fig5-energydistribution}\,(b) and (d)] are therefore determined by, respectively, the mean and the 
standard deviation of the pinning energies,  at  $T_{f}$, of the individual vortices in a given image. In particular, the displacement of the maximum of the distribution with respect to the position of the $\delta$-peak energy-distribution of a perfect vortex lattice of the same density is a measure of the mean pinning energy. As far as the vortex densities of Fig.\,~\ref{fig:Fig5-energydistribution}\, are concerned, the average $B_{int} = 0.8$\,mT yields a  $\delta$-peak-maximum at $\mathcal E_{int} = 2.5\varepsilon_{0}$. 
By comparison, the maxima  of the distributions for both investigated crystals in
Figs.\,~\ref{fig:Fig5-energydistribution}\,(a) and (b) occur at
$\mathcal E_{int}\approx 3.2\varepsilon_{0}$. The average pinning energy
per unit length is therefore $\mathcal E_{p} \sim 0.7
\varepsilon_{0}$, while the variance in pinning energy is given by
the width of the distribution, $(\langle \mathcal E_{p}^{2}\rangle
- \langle \mathcal E_{p} \rangle^{2})^{1/2}  \sim 0.5
\varepsilon_{0}$. Note that $3.2\varepsilon_{0}$ corresponds to
the interaction energy of a triangular vortex lattice with
$\Phi_{0}n_{v} =1$\,mT, {\em i.e.} the external field applied during the
experiments. This means that the average interaction energy is
determined by vortex-rich areas, with $\Phi_{0}n_{v} \gtrsim
1$\,mT. However, the vortex density also presents vortex-poor areas
so that the average $B_{int} = 0.8$\,mT. 


The large absolute values of the inferred pinning energies can be understood if one combines the notion that the crystals show local variations both of the critical temperature $T_{c} = T_{c}(\mathbf r)$ and of the line energy $\varepsilon_{0} = \varepsilon_{0}(\mathbf r,T)$, and that $T_{f}/T_{c} \lesssim 1$. As the crystal is cooled below $T_{c}$, vortices will avoid regions of higher $T_{c}$ and $\varepsilon_{0}$, and accumulate in regions with lower values of these parameters. 
They will remain trapped in such regions as the temperature is lowered below $T_{f}$. The large absolute values and variances of the pinning energies revealed by the decoration experiment are caused by the local variations of $T_{c}(\mathbf r)$, which manifest themselves through the temperature dependence of the line energy,  $\varepsilon_{0}(\mathbf r,T) = \varepsilon_{0}(\mathbf r,0)[1-T/T_{c}(\mathbf r)]$. More specifically, the width of the inferred pinning-energy distribution (Fig.~\ref{fig:Fig5-energydistribution}) should correspond to the width $\Delta \varepsilon_{0}(T_{f})$ of the line energy distribution,
\begin{equation}
0.5 \varepsilon_{0}(0)(1-T_{f}/T_{c}) \sim \Delta \varepsilon_{0}(T_{f}) .
\label{eq:width}
\end{equation}
Near to the critical temperature, $\Delta \varepsilon_{0}(T) = \varepsilon_{0}(0) T\Delta T_{c}/T_{c}^{2}$  is determined mainly by the width $\Delta T_{c}$ of the distribution of local $T_{c}(\mathbf r)$. Solving Eq.~(\ref{eq:width}) then yields $T_{f} = T_{c} / [1 + \Delta T_{c}/0.5T_{c}]$.  Taking $T_{c} = 24$ K, and estimating $\Delta T_{c} \approx 0.8$ K from the DMO data of Fig.~\ref{fig:mo}, one obtains  a freezing temperature $T_{f} =  0.94 T_{c}$ for $x = 0.075$; the same exercise with $T_{c} = 19$\,K and $\delta T_{c} = 0.5$\,K yields  and $T_{f} = 0.95 T_{c}$ for the crystal with $x = 0.1$. Thus, the analysis of the inhomogeneous and disordered vortex
distribution, as well as the vortex distribution near steps and edges, is fully consistent with the observed patterns having been
frozen between $T = 0.9$ and $0.95 T_{c}$. We can draw the same conclusion from
the local variations of the vortex density.  For example, for the crystal \#2 with $x = 0.075$, the largest local vortex gradient
correspond to $0.15$\,mT/$\mu$m or $1\times10^{8}$\,A\,m$^{-2}$. This value is consistent with the critical
current density of Ba(Fe$_{0.925}$Co$_{0.075}$)$_{2}$As$_{2}$ crystal \#2 at 23\,K.

\begin{figure}[t]
\hspace{-6mm}
\includegraphics[width=0.51\textwidth]{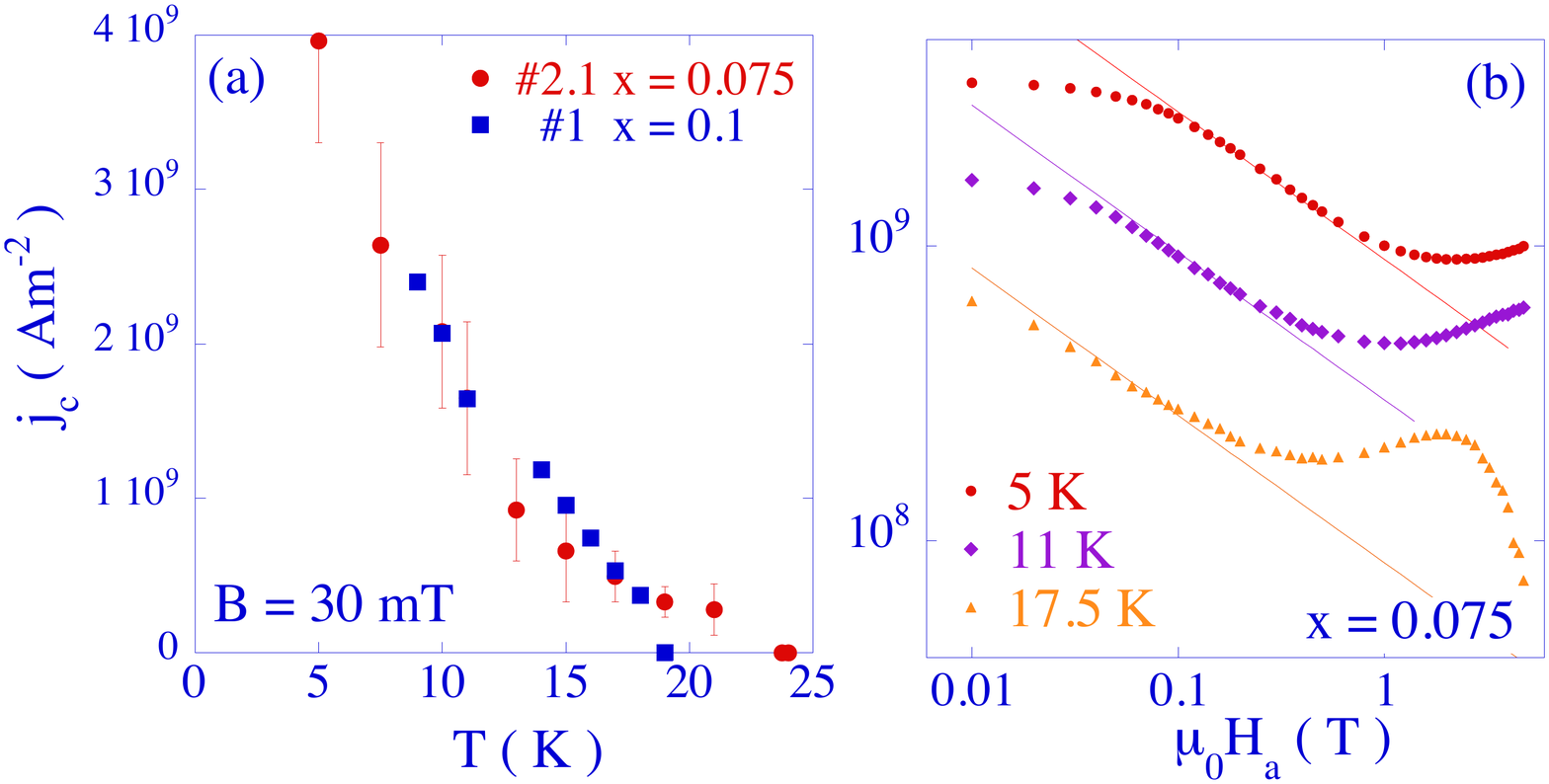}
\vspace{-20mm}
\caption{(Color online) Critical-current densities in
our Ba(Fe$_{0.925}$Co$_{0.075}$)$_{2}$As$_{2}$
crystals. (a) Temperature-dependence of the low-field $j_{c}$ of crystals \#1 ($x=0.1$)
and \#2.1 ($x=0.075$), as obtained from MOI. Error bars represent the dispersion of $j_{c}$ within a given crystal. (b) Field-dependence of $j_{c}$ for  crystal \#2.1, obtained from magnetic hysteresis measurements using a SQUID magnetometer. Straight lines indicate fits with Eq.~(\protect\ref{eq:jc(B)}), see section~\protect\ref{section:discussion}.}
\label{fig:jc}
\end{figure}

At low temperatures, the spatial variations of the magnitude of the line energy $ \varepsilon_{0}(\mathbf r , 0)$ are dominant. 
These correspond to the variations in space of  the superfluid density, \cite{Kalisky} and are responsible for the non-zero low-$T$ pinning force associated from spatial inhomogeneity.  A spatially homogeneous superfluid density would imply a vanishing
(or logarithmically weak) pinning energy  at low $T$, at odds with the existence of a large critical current density (see, \em e.g. \rm , Fig.\,~\ref{fig:jc}). 
As in all charge-doped
single-crystalline iron-based superconductors, the 
 critical current of Ba(Fe$_{1-x}$Co$_{x}$)$_{2}$As$_{2}$ 
is composed of a contribution from strong, extrinsic pins, and from a contribution
from pinning by atomic sized-point pins.  The latter dominates at high
fields [here, above 1\,T at 5\,K, and above 0.2\,T at 17.5\,K,
see Fig.\,~\ref{fig:jc}(b)],\cite{Kees}  while the former  contribution manifests itself
as a low-field plateau \cite{Kees1,vdBeek2002}
\begin{equation}
j_{c}  =  \pi^{1/2} \frac{f_{p}}{\Phi_{0}\varepsilon_{\lambda}}
\left( \frac{U_{p} n_{i}}{\bar{\varepsilon}_{0}}\right)^{1/2} \hspace{1cm}(B \ll B^{*})
\end{equation}
followed by a power-law  in
the flux density $B$,\cite{vdBeek2002,Kees1}
\begin{equation}
j_{c}(B)  =  \frac{f_p}{\Phi_{0}\varepsilon_{\lambda}} \left( \frac{U_{p}n_{i}}{\varepsilon_{0}}\right)\left(\frac{\Phi_{0}}{B}\right)^{1/2} \hspace{1cm}(B \gg B^{*}).
\label{eq:jc(B)}
\end{equation}
The crossover field $B^{*}$  is that above which the number of effective pins per
vortex is limited by the intervortex repulsion, $f_{p}$ is the maximum pinning force exerted by a single strong
 pin, $n_{i}$ is the pin density, and $U_{p}/[\mathrm J]$
 is the pinning energy gained by a vortex line traversing such a pin.
 The measurement of the low-field critical current density $j_{c}(0)$ and
 the slope $\partial j_{c}(B)/\partial B^{-1/2}$ allows one to
 eliminate $n_{i}$ and to obtain $f_{p} = \pi \Phi_{0}^{3/2}\varepsilon_{\lambda} \left\{ j_{c}^{2}(0)/[\partial j_{c}(B)/\partial B^{-1/2}]\right\}$
 from  experimental data without further assumptions. We find that, at 5 K,
  $f_{p} \approx 3 \times 10^{-13}$\,N for both Ba(Fe$_{0.925}$Co$_{0.075}$)$_{2}$As$_{2}$ crystal \#2.1 and
Ba(Fe$_{0.9}$Co$_{0.1}$)$_{2}$As$_{2}$ crystal \#1.

The identification of the strong pins with 
regions of lower $\varepsilon_{0}(T)$ means that $f_{p}$ should be interpreted in terms of the local
maxima of the position-dependent force  $f(\mathbf r) =
\int_{\delta z} \nabla  \varepsilon_{0}  (\mathbf r) d z$
experienced by  vortices as they move through the sample. Here
$\delta z$ is the maximum extent of a region of low $\varepsilon_{0}(T)$ 
along the field direction.  We approximate 
\begin{equation}
f_{p} \sim \Delta \varepsilon_{0} \left(\frac{\delta z}{\delta \varrho} \right),
\label{eq:fp}
\end{equation}
where $\delta \varrho$ is the length scale characterizing the disorder in the direction perpendicular to the field, and $\Delta \varepsilon_{0}$ is the standard deviation of the $\varepsilon_{0}(\mathbf r)$ distribution in the crystal. The pinning energy $U_{p}
\sim f_{p}\delta \varrho$.  A comparison of Eq.~(\ref{eq:fp}) with the value of $f_{p}$ obtained from $j_{c}$ yields $\Delta \varepsilon_{0} \sim 3\times 10^{-13}$ J\,m$^{-1}$ for a unit aspect ratio $\delta z/\delta \varrho$.

In a next step, we evaluate the ratio of $f_{p} / \bar{|\mathbf f_{i}|}$ to
 obtain the average distance between effective pins,
  $\bar {\mathcal{L}} = 60$ nm. Using Eq.~(17) of Ref.~\onlinecite{vdBeek2002},
  which has $\bar{\mathcal{L}} = ( \varepsilon_{1}/ \pi n_{i}U_{p})^{1/2}$, one
   finds  $(n_{i}\delta z)^{-1/2} \sim 60$\,nm. With all parameters known,
 the low-field value of the critical current density is reproduced as 
 \begin{equation}
     j_{c}     \approx  \pi^{1/2} \frac{ \Delta \varepsilon_{0}}{\Phi_{0}\varepsilon_{\lambda}} \sqrt{ n_{i} \delta z} \frac{\delta z}{\delta \varrho}  \sqrt{ \frac{ \Delta \varepsilon_{0} }{ \bar{\varepsilon}_{0}}}  = 8\times10^{9}\,\mathrm{A\,m^{-2}},
\end{equation}
in fair agreement with the data of
Fig.\,~\ref{fig:jc}\,(a). The investigated features of vortex
pinning in  Ba(Fe$_{1-x}$Co$_{x}$)$_{2}$As$_{2}$, including the
disordered vortex patterns and the critical current density, are
therefore consistently described by the presence of spatial
variations of the superfluid density on the scale of \em several dozen \rm nanometers, in agreement
with the conjecture of Ref.~[\onlinecite{Yamamoto}].

Note that the observed spatial structures at the macroscopic
(Fig.\,~\ref{fig:mo}) and mesoscopic (Fig.\,~\ref{fig:deco})
levels are not those responsible for the critical current. The
random vortex positions observed in the decoration experiments are
 determined by the underlying nanoscale disorder, an
observation consistent with the fact that disordered vortex
structures have been observed up to high
fields.\cite{Eskildsen,Inosov}

One may speculate about the possible link between the existence of nm-sized regions of reduced superfluid density, the local variation of the dopant atom density, and the effect of the overall doping level. For instance, one would expect the fluctuations
of the Co density to be more important at lower doping levels, yielding larger local fluctuations of $\varepsilon_{0}$. However,
 given the much larger values of the penetration depth at low doping, we have
 not been successful in performing Bitter decorations on the relevant crystals.
Recent STS studies have reported substantial variations of the value of
  the superconducting gap on a scale of 10 to 20\,nm.\cite{Yi Yin,Massee2009}
   These local variations of the gap magnitude should correspond to the variations
    of the critical temperature and therefore lead to vortex pinning. Although it is
    tempting to relate our results to the nanoscale disorder observed in the STS
    gap-maps, it should be remarked that the spatial scale of the variations in the
     gap maps is a factor of 3-6 smaller than that found from the analysis of the
     data presented here. This would correspond to a concomitantly  larger $j_{c}$
     in the samples used in Refs.~[\onlinecite{Yi Yin,Massee2009}].
\\

 \section{Conclusion}

Bitter-decoration imaging of the disordered vortex distribution in
superconducting Ba(Fe$_{1-x}$Co$_{x}$)$_{2}$As$_{2}$
single-crystals with $x=0.075$ and $x=0.1$ reveals a substantial
local variation of pinning energies and pinning forces. The
magnitude of these fluctuations is suggested to stem from
nanoscale spatial variations of $T_{c}$ and/or the superfluid
density due to an inhomogeneous distribution of dopant atoms. The
spatial scale of the variations is inferred from the correlation
of the features of the vortex distributions with global and local
critical current density measurements. The macroscopic spatial
variations of the critical temperature observed using
magneto-optical imaging give an idea of the magnitude of the
$T_{c}$ variations in the crystals, but are unrelated to the
measured pinning properties. The same can be said for mesoscopic
disorder structures observed by single-vortex imaging. An
important corollary of our work is the fact that the observed
vortex distributions are frozen, at a length scale of the lattice
spacing, at a high temperature close to $T_{c}$.

\section*{Acknowledgements} We acknowledge J. J. Z\'{a}rate for his support on the profilometer measurements. 
This work was made possible due to the support of the ECOS-Sud-MINCyt France-Argentina bilateral program, Grant
\#A09E03. Work done in France was partially funded by the grant  ``MagCorPnic'' of the R\'{e}seau Th\'{e}matique de Recherche Avanc\'{e}e   ``Triangle de la Physique'' du Plateau de Saclay, and by the Agence Nationale de la Recherche grant "PNICTIDES". Work done in Bariloche   was partially funded by PICT 2007-00890 and PICT 2010-294 associated to  PRH74 program on ``Nanoscience and Nanotechnology'' from the   ANPCyT, and from the Argentinean Atomic National Commission  (CNEA). N.R.C.B. holds a doctoral scholarship from Conicet \&   ANPCyT.


\begin{thebibliography}{99}
\bibitem{check} We have checked our summing method by calculating the interaction energy distribution of a perfect
Abrikosov lattice, which should be a $\delta$-function. After neglecting the contribution from the vortices situated less than
$4\lambda_{ab}$ from the edges of the image, we obtain, for the perfect lattice, a well-defined $\delta$-peak for the interaction
energy distribution.
\end{thebibliography}
\end{document}